\documentclass[prd,a4paper,
amsmath, amssymb, floatfix,
twocolumn,nofootinbib,wrapfloat byrevtex]{revtex4}
\usepackage[english]{babel}
 \usepackage{amsmath,mathrsfs,upgreek,graphicx,placeins,float,hyperref,tikz}
 \usepackage{color}
 \usepackage{booktabs}
 \usepackage{soul}
 \usepackage{multirow}
\usepackage{appendix}

\newcommand{\bea}[1]{\begin{align}\label{#1}}
\newcommand{\eea}{\end{align}}
\newcommand{\beq}[1]{\begin{equation}\label{#1}}
\newcommand{\eeq}{\end{equation}}
\newcommand{\rmd}{\mathrm{d}}
\newcommand{\bfq}{\mathbf{q}}
\newcommand{\bfb}{\mathbf{b}}

\renewcommand{\eqref}[1]{(\ref{#1})}
\newcommand{\figref}[1]{\text{Figure}~\ref{#1}}
\newcommand{\secref}[1]{\text{Section}~\ref{#1}}
\newcommand{\mca}[1]{\mathcal{#1}}
\newcommand{\mrm}[1]{\mathrm{#1}}
\newcommand{\mbf}[1]{\mathbf{#1}}
\newcommand{\ddq}{\int \frac{\mathrm{d}^d q}{(2\pi)^d}}
\newcommand{\ddl}{\int \frac{\mathrm{d}^d{\ell}}{(2\pi)^d}}

\numberwithin{equation}{section}

\begin{document}

\title{Gravitational Deflection of Vector Photons via Effective Field Theory}

\author{Yihan Ma \& Ding-fang Zeng}
\email{yihan.ma@emails.bjut.edu.cn,dfzeng@bjut.edu.cn}
\affiliation{School of Physics and Optoelectronic Engineering, Beijing University of Technology}
\date{\today}

\begin{abstract}
Gravitational scattering of the electromagnetic field from a heavy scalar field provides a fundamental testbed for understanding the deflection of light by massive bodies. 
In many approaches based on effective field theory, the calculation of scattering amplitudes quickly becomes complicated due to the large number of Feynman integrals required, especially beyond leading order. 
In this work, we study this problem using effective field theory in the weak field approximation. 
We utilize Integration-By-Parts reduction techniques to precisely examine the long-range contributions governed by terms in the amplitude which are non-analytic in momentum transfer. 
Using geometric optics and the eikonal approximation, we derive expressions for the deflection angle and find the origin of differences relative to earlier works.
\end{abstract}

\maketitle

\section{Introduction}

The shadowy images of supermassive black holes at the center of the M87 galaxy and the Milky Way, obtained by the Event Horizon Telescope Collaboration \cite{EHT-2019.04-1, EHT-2019.04-2, EHT-2019.04-3,EHT-2022}, have gained widespread attention in recent years. 
The theoretical basis of this achievement is closely tied to the lensing formula derived by K.S. Virbhadra and G. F.R. Ellis, which describes the deflection angle of light rays passing through the gravitational field of a Schwarzschild black hole \cite{Virbhadra-2000.09, Gibbons-2008}.
Accurately calculating gravitational lensing effects is essential for obtaining clear images of black holes, determining how well we can understand them and their environment.

Based on the work of K.S. Virbhadra and G.F.R. Ellis, S. Frittelli et al. \cite{Frittelli-2000.02}, V. Bozza et al. \cite{Bozza-2001.09}, and N. Tsukamoto \cite{Tsukamoto-2017.03} refined the methodology for the calculation of the gravitational lensing effects in strong gravitational fields. 
Similar to the case in the weak field regime, alternative methods are available for calculating the deflection angle, such as the impulse approximation and topological methods based on the Gauss–Bonnet theorem \cite{Gibbons-2008, Heidari-2024}.
Besides these approaches, Effective Field Theory (EFT) \cite{Donoghue-1994.05, Donoghue-1994.09, Donoghue-1999.06} has also been introduced in recent years to calculate the deflection angle, as a special case of the more general phenomena of gravitational scattering.

In conventional applications of this method, the deflection angle of light is derived from the photon–black hole scattering amplitude \cite{Bai-2017.03,Guadagnini-2002,Brunello-2022}. 
However, the complexity of the vertices for graviton–photon interactions and graviton self-interactions increases significantly with increasing order in perturbation theory. 
To address this challenge, modern EFT approaches employ the Kawai–Lewellen–Tye (KLT) relation \cite{Bern-2002, Bern-2008.10}, which expresses the gravitational scattering amplitude as a product of Yang–Mills amplitudes. 
Combining with the unitarity techniques \cite{Dixon-2014}, this approach enables the direct determination of terms non-analytic in the momentum transfer at the one-loop level. 
After Fourier transformation, these terms yield long-range gravitational effective potentials \cite{Bohr-2015.02, Bohr-2016}, from which the photon deflection angle can be systematically derived.

Another way of reducing the workload of the EFT method is to employ modern loop integration techniques, such as Integration-by-Parts (IBP) reduction. 
This technique allows us to focus on very few master integrals instead of directly calculating a vast number of Feynman integrals. 
A major advantage of this technique is its applicability to higher-loop calculations. 
In ref.\cite{Brunello-2022}, this method was applied to calculate the deflection angle of scalar photon trajectories at the one-loop level. 
In the present work, we extend this approach to one-loop amplitudes for the more realistic scenario of vector-photon scattering from black holes. 
We will take a heavy scalar field as a proxy for a black hole. 
Most of the repetitive and routine aspects of the calculation, such as generating Feynman rules and reduction of Feynman integrals, will be performed by computer algebra systems. 
Several packages are available for IBP reduction, such as FIRE6 and Kira (both are based on the Laporta algorithm \cite{Laporta-2000}), and LiteRed \cite{Lee-2012, Lee-2014}. 
For this work, we chose LiteRed. 

With the black hole represented by a massive scalar field $\phi$, the photon represented by the standard vector field $A$ and the graviton described by a second-order tensor field $h$, we write the effective action controlling the interaction between these objects as follows:
\beq{din-effact}
\mrm{e}^{iS_{\mrm{eff}}}{=}\int \mca{D}[\phi,A,h,\bar{c}_{h},c_h,\bar{c}_A,c_A]\mrm{e}^{i(S_{\phi}+S_{h}+S_{A}+S_{\mrm{ghost}})}.
\eeq
$c_A\&c_h$ here are the ghost fields of the electromagnetic and gravitational fields, respectively.
\begin{align}
S_{\phi} &= \int \rmd^4 x \frac{\sqrt{-g}}{2}(\partial^{\mu}\phi \partial_{\mu}\phi- m^2 \phi^2),
\label{phi-act}
\\
S_{h} &= \kappa^{-2} \int \rmd^4 x \sqrt{-g}\left(-2R + \kappa^2 \mca{G}_{h}^{\mu}\mca{G}_{h\mu}\right) ,
\label{h-act}
\\
S_A &= \int \rmd^4 x \sqrt{-g}
\big(-\frac{1}{4} F^{\mu\nu}F_{\mu\nu} + \frac{1}{2} \mca{G}_{A}^2 \big),
\label{A-act}
\end{align} 
where $\kappa=\sqrt{32\pi G_N}$, while $\mca{G}_{\mrm{h}}^{\mu}$, $\mca{G}_{\mrm{A}}$ are the gauge-fixing functions of the gravitational and electromagnetic fields,
\begin{equation}\label{gaugeFix}
\mca{G}_{h}^{\mu} = \Gamma^{\mu}
~,~\mca{G}_{A} = \partial^{\mu} A_{\mu}.
\end{equation}
Using the xAct package\cite{Garcia-2007.10, Garcia-2008.10-1, Garcia-2008.10-2}, we can derive the desired Feynman rules from \eqref{phi-act}-\eqref{A-act}. The detailed expressions have been listed in the Appendix A.
We have confirmed that they are consistent with those provided by B. Latosh in FeynGrav \cite{Latosh-2022, Latosh-2023.11}.

The structure of this paper is as follows. 
In \secref{chpt:amp}, we calculate the amplitude for photon scattering from a heavy scalar field in the center-of-mass frame and present the results in the small-angle approximation. 
In \secref{chpt:ang}, we derive the deflection angle of photon trajectories passing through the gravitational field of massive bodies using the eikonal and geometric optic approximations respectively, comparing our results with those in the existing literature. 
Finally, in \secref{chpt:con}, we summarize the results of our calculation and briefly discuss potential questions.

\section{Calculation of scattering amplitudes}\label{chpt:amp}

\begin{figure}[htbp]
\includegraphics[totalheight=30mm]{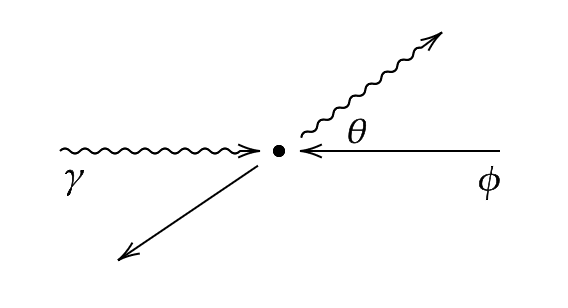}
\caption{Diagram of $\gamma-\phi$ scattering in the center-of-mass frame.}
\label{fig:sanshe}
\end{figure}

We choose to calculate the tree-level and one-loop level amplitude of photon-black hole scattering in the center of mass frame, as shown in \figref{fig:sanshe}. 
We take the momentum direction of the incident photon as that of the positive z-axis. 
Thus, in the in-out formalism, the polarization vectors of the ingoing and outgoing photons are as follows:
\begin{align}
&\hspace{-0mm}\epsilon_{\mrm{i}}^{+\mu}{=}\frac{1}{\sqrt{2}}(0,1,i,0),
\notag
\\
&\hspace{-0mm}\epsilon_{\mrm{i}}^{-\mu}{=}\frac{1}{\sqrt{2}}(0,1,-i,0),
\notag
\\
&\epsilon_{\mrm{o}}^{+\mu}{=}\frac{1}{\sqrt{2}}(0,\cos\theta_{\mathrm{cm}},i,-\sin\theta_\mathrm{cm}),
\notag
\\
&\epsilon_{\mrm{o}}^{-\mu}{=}\frac{1}{\sqrt{2}}(0,\cos\theta_\mathrm{cm},-i,-\sin\theta_\mathrm{cm}).
\label{epo}
\end{align}
The superscript ``+'' here denotes the right-handed polarization, while the ``-'' supscript denotes the left-handed polarization.

According to the on-shell condition of external particles as well as the conservation of four-component momentum, we have
\begin{align}
&k_{\mrm{i}}^2=k_{\mrm{o}}^2=0~,~p_{\mrm{i}}^2=p_{\mrm{o}}^2=m^2,
\label{205}
\\
&k_{\mrm{i}}+p_{\mrm{i}}-k_{\mrm{o}}-p_{\mrm{o}}=0 \label{206},
\end{align}
where $k_{\mrm{i}}\&k_{\mrm{o}}$ represent the incident and outgoing photons respectively, $\omega$ is their energy; $p_{\mrm{i}}\&p_{\mrm{o}}$ represent the initial and final momentum of the heavy scalar field, with $m$ being its invariant mass. 
The Mandelstam variables associated with the scattering process are:
\begin{equation}\label{207}
s=(k_{\mrm{i}}+p_{\mrm{i}})^2~,~t=(k_{\mrm{i}}-k_{\mrm{o}})^2~,~ u=(k_{\mrm{i}}-p_{\mrm{o}})^2.
\end{equation}
As usual, the three Mandelstam variables satisfy the standard relation
\begin{equation}\label{208}
s+t+u=2m^2.
\end{equation}
Via these definitions, we can easily show that in the low-energy regime, the following expressions hold,
\begin{equation}\label{man-aprx}
s\simeq{}m^2+2m\omega~,~t\simeq{}-\bfq^2~,~u\simeq{}m^2-2m\omega+\bfq^2,
\end{equation}
where $q$ is the transferred momentum four-vector, with $\bfq$ being its three-dimensional component. 
Using Mandelstam variables, we can rewrite the scattering angle $\theta_{\mathrm{cm}}$ in terms of the polarization vectors,
\begin{equation}
\begin{aligned}
\cos\theta_{\mathrm{cm}}&=1+\frac{2st}{(s-m^2)^2},\\
\sin\theta_{\mathrm{cm}}&=\frac{2[st(su-m^4)]^{1/2}}{(s-m^2)^2}.
\end{aligned}
\end{equation}

\subsection{Scattering amplitude at tree-level}

\begin{figure}[htbp]
\centering{
\includegraphics[totalheight=30mm]{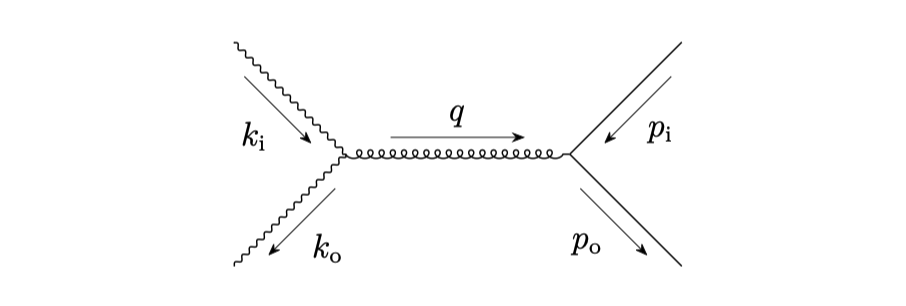}
}
\caption{Tree-level Feynman diagram of t-channel}
\label{fig:tree}	
\end{figure}

At the tree level, only the t-channel diagram contributes to the amplitude of photon-black hole scattering, as shown in \figref{fig:tree}, where $ q^{\mu} = k_{\text{o}}^{\mu} - k_{\text{i}}^{\mu} = (0, \mathbf{q}) $ represents the momentum transfer during the process.  

According to the Feynman rules listed in the Appendix A,  in the case the polarization states of the the incident and scattered photons are identical, the scattering amplitude at the tree level can be written as follows:
\begin{align}
&i\mca{M}^{(\mrm{t})}= \epsilon^{+}_{\mrm{i}\mu}(\epsilon^+_{\mrm{o}\nu})^* i \mca{M^{(\mrm{t})\mu\nu}} 
\label{tre-polsma-0}
\\
&\hspace{9mm}=i\kappa^2\omega^2-\frac{1}{4}i\kappa^2m^2-\frac{i\kappa^2m^2\omega^2}{t}
\label{tre-polsma-1}
\\
&\hspace{9mm}\simeq{}-\frac{i\kappa^2m^2\omega^2}{t}.
\label{tre-polsma-2}
\end{align}
The weak deflection limit implies that $t\ll\omega^2\ll{}m^2 $. 
Considering this fact, we performed a double series expansion of the scattering amplitude above. 
The expansion in $t$ is retained up to $ \mathcal{O}(t^0) $, while the expansion in $\omega$ is preserved up to $ \mathcal{O}(\omega^4)$. 
Due to the condition $t\ll\omega^2\ll{}m^2 $, only the third term in eq\ref{tre-polsma-1} is significant.

In the case the polarization states of the incident and scattered photons differ, we have
\begin{align}
&\epsilon^{+}_{\mrm{i}\mu}(\epsilon^-_{\mrm{o}\nu})^*\mca{M^{\mu\nu}}_{\mrm{t}} 
=\i \kappa ^2 \omega ^2+\frac{i\kappa^2}{2}m\omega.
\label{tre-poldiff}
\end{align}
We will derive a local effective potential after Fourier transformation of the amplitude above. 
In the weak deflection limit, we will focus on the long range interaction only, which decays as $r^{-n}$. 
Therefore, all terms corresponding to local interactions will be ignored or neglected.
\begin{figure*}[htbp]
\includegraphics[totalheight=160mm]{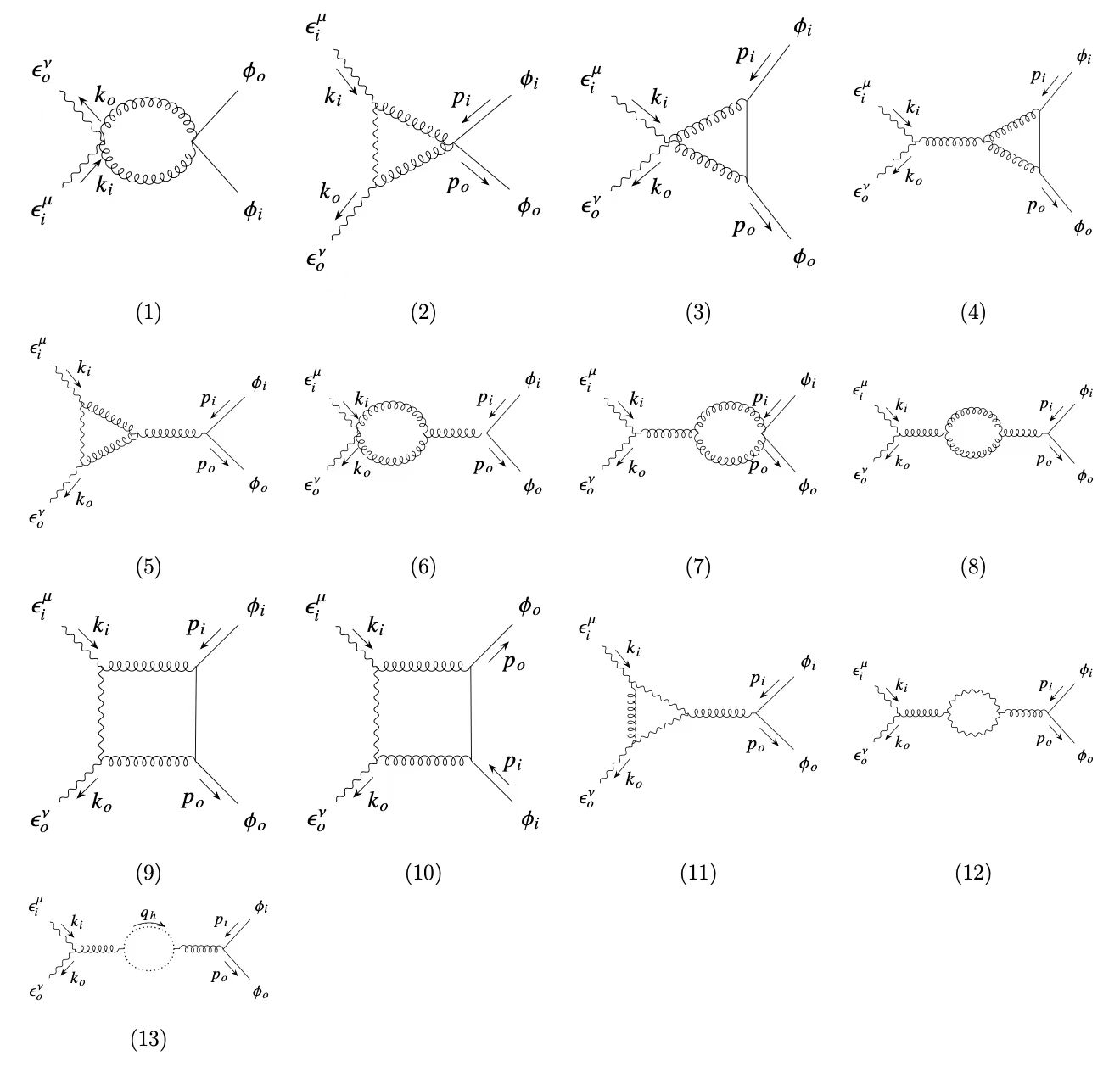}
\caption{The one-loop Feynman graphs of the scattering of a photon from a heavy scalar field which need to be calculated}
\label{figOneLoopDiagrams}
\end{figure*}

\subsection{Scattering amplitude at one-loop level}

In our calculations, we treat the gauge field $A_{\mu}$ and gravitational field $h_{\mu\nu}$ according to the classical Feynman diagram method, following refs.\cite{Bai-2017.03,Brunello-2022}. 
However, we choose gauge fixing conditions in such a way that only the gravi-ghost contributes to the scattering amplitude at the one loop level. 
At this level, the diagrams that need to be calculated are shown in \figref{figOneLoopDiagrams}, where we used $\ell_h$ to represent the momentum of ghost field in the last diagram.  
At the same time, in order to simplify the expressions, we will introduce the following notations to write the scattering amplitude:
\begin{align}
D_1&\equiv\ell^2~,~D_2\equiv(k_{\mrm{i}}-{\ell})^2~,~D_3\equiv(k_{\mrm{i}}-k_{\mrm{o}}-{\ell})^2,
\notag
\\
D_4&\equiv(p_{\mrm{i}}+{\ell})^2-m^2~,~D_5\equiv({\ell}-p_{\mrm{o}})^2-m^2.
\label{equ:simprop}
\end{align}
The first issue we encounter when calculating the one-loop diagram is the treatment of $\epsilon_{\mrm{i},\mrm{o}} {\cdot}{\ell}$ factors in the amplitude denoted by the diagram (1) in Figure \ref{figOneLoopDiagrams},
\begin{align}
&\epsilon_{\mrm{i}\mu}\epsilon^*_{\mrm{o}\nu}V^{\mu\nu\mu_1\nu_1}_{AAhh}[k_\mrm{i},{-}k_\mrm{o},{-}{\ell},{-}(k_\mrm{i}-k_\mrm{o})] 
\label{cuppolver}
\\
&\sim \ddl\epsilon_{\mrm{i}\mu}\epsilon^*_{\mrm{o}\nu}\frac{{\ell}^{\mu}{\ell}^{\nu}}{D_2 D_3}.
\notag
\end{align}
This integration over $q$ is intractable due to the appearance of the scalar product $q\cdot\epsilon$ in the numerator of the integral. 
To address this issue, we perform tensor decompositions of the loop integral,
\begin{align}
&\ddl \frac{{\ell}^{\mu}{\ell}^{\nu}}{D_2 D_3}
\label{sam-tid}
\\
&
{=}
{-}\ddl\frac{k^{\mu}_{\mrm{i}}k^{\nu}_{\mrm{o}}}{D_2 D_3}
{-}\ddl\frac{k^{\nu}_{\mrm{i}}k^{\mu}_{\mrm{o}}}{D_2 D_3}.
\notag
\end{align}
This is just one example of a tensor decomposition. 
Because of the large number of terms involving $\epsilon{\cdot}{\ell}$ in the loop integrals, many of which are quite complicated, such as $\frac{\epsilon{\cdot}{\ell}~\epsilon^* {\cdot}{\ell}}{D_1 D_2 D_3 D_4}$,  this tensor decomposition is extremely tedious. 
We use FeynCalc \cite{Mertig-1991, Shtabovenko-2020.11} to perform this work.
After the tensor decomposition, we proceed to the calculation of loop integrals. 
Unlike the traditional Feynman diagram methods used in \cite{Bai-2017.03} and the on-shell techniques employed in \cite{Bohr-2016}, we choose to perform IBP reduction in the loop integrals. 
Consider the following integral, where $\mathcal{I}$  denotes a general scalar integrand,
\begin{equation}\label{ibp-1}
I=\int \prod_{i=1}^L \rmd^d {\ell}_i \mca{I},
\end{equation}
after an infinitesimal transformation of the loop momentum ${\ell}_i^{\mu} \to {\ell}_i^{\mu} + a_{ij} p_j^{\mu}$, we obtain
\begin{align}
&\int \prod_{i=1}^L \rmd^d {\ell}_i \mathcal{I}
\label{ibp-2}
\\
&{\to} {\int} \prod_{i=1}^L \rmd^d {\ell}_i\left( 1{+}a_{ij}\frac{\partial p^{\mu}_j}{\partial{\ell}^{\mu}_i} \right) \left(1{+}a_{ij}p^{\mu}_{j}\frac{\partial}{\partial{\ell}^{\mu}_i} \right)\mathcal{I},
\notag
\end{align}
where $p^{\mu}_j \in \{{\ell}_1,...,{\ell}_L,k_1,...,k_{n+1} \} $, with $k$ representing the external momentum. 
In dimensional regularization, this transformation does not change the value of these integrals, so we have
\beq{ibp-3}
0=\int \prod_{i=1}^L \rmd^d {\ell}_i~a_{ij} \frac{\partial }{\partial{\ell}^{\mu}_i}\left( p^{\mu}_j\mathcal{I}\right).
\eeq
Since the integrand of eq.\eqref{ibp-1} is a scalar, it is invariant under the Lorentz transformation,
\beq{ibp-5}
I \to I (k_1+\delta k_1,...) = \left( 1+ \sum_i \omega^{\mu}_{\nu} k^{\nu}_{i} \frac{\partial }{\partial k^{\mu}}\right) I,
\eeq
which means that,
\begin{align}
&\left( \sum_i \omega^{\mu}_{\nu} p^{\nu}_{i,\mrm{ext}} \frac{\partial }{\partial k^{\mu}}\right) I=0 \label{ibp-6}.
\end{align}
By exchanging the indices $\mu\&\nu$, and applying $\omega^{\mu}_{\nu}=-\omega^{\nu}_{\mu}$, we will obtain,
\begin{align}
&\sum_i \left( p^{\nu}_{i,\mrm{ext}} \frac{\partial }{\partial k^{\mu}}- p^{\mu}_{i,\mrm{ext}} \frac{\partial }{\partial k^{\nu}}\right) I
=0.
\label{ibp-8}
\end{align}
Using relations between the Feynman integrals provided by eq.\eqref{ibp-3} and eq.\eqref{ibp-8}, we can express the integral $I$ as a linear combination of simpler scalar integrals
\begin{equation}\label{lincomb}
I= \sum_i c_i I_i,
\end{equation}
where $c_i$s refer to coefficients depending on the Lorentz invariants, and $I_i$ refers to integrals that cannot be simplified further through IBP reduction, i.e. the master integrals. 
There are many software packages available for the IBP reduction of Feynman integrals, in this work we choose LiteRed.

We take the Lorentz invariants in eq.\eqref{equ:simprop} as a basis, expressing all scalar products of external and loop momentum as linear combinations of this basis. 
For example
\begin{equation}\label{bascho}
k_{\mrm{i}}{\cdot}{\ell} = -\frac{D_2-D_1}{2}~,~p_{\mrm{i}}{\cdot}{\ell} = \frac{D_4-D_1}{2}~,~... .
\end{equation}
For convenience, we will use the following notations to denote general Feynman integrations
\beq{simint}
I_{[n_1,n_2,n_3,n_4]}= \ddl \frac{1}{D_1^{n_1}D_2^{n_2}D_3^{n_3}D_4^{n_4}}.
\eeq
As an example to demonstrate the IBP reduction techniques in the calculation of loop integrals, consider the following integral,
\begin{align}
&\ddl \frac{k_{\mrm{i}}{\cdot}{\ell}~p_{\mrm{i}}{\cdot}{\ell}}{D_1 D_2 D_3 D_4}
\label{sam:ibp-3}
\\
&=-\frac{1}{4}\ddl \frac{D_2 D_4+D_1^2-D_1D_2-D_1 D_4}{D_1 D_2 D_3 D_4} 
\notag
\\
&\equiv{-}\frac{1}{4}\left(I_{[1,0,1,0]}{+}I_{[-1,1,1,1]}{-}I_{[0,0,1,1]}{-}I_{[0,1,1,0]}  \right) 
\label{sam:ibp-4}
\end{align}
\begin{align}
&={-}\frac{1}{4}\left[ I_{[1,0,1,0]}{+}\frac{I_{[0,1,0,1]}\left(m^4\epsilon{-}2m^2s\epsilon{+}s^2\epsilon{+}st\right)}{\epsilon \left(m^2{-}s\right)^2}\right. 
\notag
\\
&\left.{-}\frac{t(\epsilon{-}1) \left(m^2{+}s\right)I_{[0,0,0,1]}}{2 m^2\epsilon(2\epsilon {-}1) \left(m^2{-}s\right)^2}-\frac{(\epsilon{-}1)I_{[0,0,0,1]}}{m^2(2\epsilon{-}1)}{-}0 \right],
\label{sam:ibp-5}
\end{align}
where the spacetime dimension is set $d = 4 - 2\epsilon$. 
After this operation, the integral we intend to compute, eq.\eqref{sam:ibp-3}, is transformed into three simpler scalar integrals $I_{[0,0,0,1]}, I_{[1,0,1,0]}, I_{[0,1,0,1]}$ as shown in eq.\eqref{sam:ibp-5}. 
It should be noted that the diagram (10) in Figure \ref{figOneLoopDiagrams} has a u-topology, while the others have s-topologies. 
The basis for these two types of diagrams are different. 
For the former, the basis is $\{D_1, D_2, D_3, D_5\}$. 
While for the latter, it is $\{D_1, D_2, D_3, D_4\}$. 
The diagram (10) should be reduced separately.

After the IBP reduction, all the loop integrals involved in the Feynman diagram of Figure \ref{figOneLoopDiagrams}  can be written as a linear combination of four master integrals. 
The corresponding amplitudes read
\begin{align}
\mathcal{M}^{(1)}&=\kappa^4m^2\omega^2I_{[1,0,1,0]},
\label{ampDiagrm1}
\\
\mathcal{M}^{(2)}&=\frac{\kappa^4m^2\omega^2}{8}I_{[1,0,1,0]},
\\
\mathcal{M}^{(3)}&=-\frac{\kappa^4m^2\omega^2}{2}I_{[1,0,1,0]},
\\
\mathcal{M}^{(4)}&=\frac{\kappa^4\omega^2}{32}(19m^2I_{[1,0,1,0]}+2m^4I_{1,0,1,1}),
\\
\mathcal{M}^{(5)}&=-\frac{\kappa^4m^2\omega^2}{24}I_{[1,0,1,0]},
\\
\mathcal{M}^{(6)}&=-\frac{11\kappa^4m^2\omega^2}{24}I_{[1,0,1,0]},
\\
\mathcal{M}^{(7)}&=-\frac{3\kappa^4m^2\omega^2}{4}I_{[1,0,1,0]},
\\
\mathcal{M}^{(8)}&=-\frac{29\kappa^4m^2\omega^2}{240}I_{[1,0,1,0]},
\\
\mathcal{M}^{(9)}&=\kappa^4\Big(\frac{7}{16}\kappa^4m^3\omega{-}\frac{\kappa^4m^3\omega}{16\epsilon}{+}\frac{5}{16}\kappa^4m^2\omega^2
\\
&{+}\frac{\kappa^4m^2\omega^2}{16\epsilon}\Big)I_{[1,0,1,0]}{-}\kappa^4\frac{1}{2}m^4\omega^2I_{[1,0,1,1]}
\notag\\
&{+}\kappa^{4}m^4\omega^4I^{(s)}_{[1,1,1,1]},
\notag
\\
\mathcal{M}^{(10)}&=\kappa^4\Big({-}\frac{7}{16}m^3\omega{+}\frac{m^3\omega}{16\epsilon}{-}\frac{35}{16}m^2\omega^2
\\
&{+}\frac{11m^2\omega^2}{16\epsilon}\Big)I_{[1,0,1,0]}{-}\frac{1}{2}\kappa^4m^4\omega^2I_{[1,0,1,1]}
\notag
\\
&{+}\kappa^{4}m^4\omega^4I^{(u)}_{[1,1,1,1]},
\notag
\\
\mathcal{M}^{(11)}&=\frac{\kappa^4m^2\omega^2}{4}\frac{\epsilon+1}{\epsilon}I_{[1,0,1,0]},
\\
\mathcal{M}^{(12)}&=\frac{9\kappa^4m^2\omega^2}{20}I_{[1,0,1,0]},
\\
\mathcal{M}^{(13)}&=\frac{11\kappa^4m^2\omega^2}{120}I_{[1,0,1,0]}.
\label{ampDiagrm13}
\end{align}
Expressions of the four master integrals can be found in \cite{R.Keith-Ellis-2008}, or computed with Package-X \cite{Patel-2017.09}. 
For readers' convenience, we have compiled them in the Appendix B. 
With these expressions, we can write the complete one-loop amplitude as explicit functions of the Mandelstam variable $s$, $t$ and $u$. 
The long-range effective potentials between the photon and black holes decay as $r^{-n}$. 
They manifest as terms non-analytic in $t$ within the one-loop amplitude, 
\begin{align}
\mca{M}_{\mrm{non-ana}}(t)&=C_1\log\big({-}\frac{\bar{\mu}^2}{t}\big)+C_2\log\big({-}\frac{m^2}{t}\big)
\label{non-ana}
\\
&+C_3 \sqrt{{-}\frac{1}{t}}+C_4 \log^2\big({-}\frac{\bar{\mu}^2}{t}\big)
\notag
\end{align}
where $\bar{\mu}^2=4\pi\mathrm{e}^{-\gamma_E}\mu^2$. 
Using eqs.\eqref{mi-1}-\eqref{itegralU1111}, we can extract all the non-analytic terms from the amplitude \eqref{ampDiagrm1}-\eqref{ampDiagrm1}. 

The weak deflection limit implies that $t\ll\omega^2\ll{}m^2 $. 
Considering this fact, we perform a double series expansion of the amplitude after the loop integration.  
Since only the non-analytic terms in $t$ are relevant, the expansion in $t$ will be retained up to $ \mathcal{O}(t^0) $, while the expansion in $\omega$ will be kept up to $ \mathcal{O}(\omega^4) $. 
The detailed expression is complicated but nonetheless comprehensible, 
\begin{align}
&i\mca{M}_{\mrm{1-loop}}=\kappa^4m^2\omega^2\Big(-\frac{253}{96}+\frac{1}{\epsilon}\Big)I_{[1,0,1,0]}
\label{loop-th}
\\
&\hspace{2.5mm}
+\kappa^4m^4\omega^2I_{[1,0,1,1]}+\kappa^4m^4\omega^4\Big(I^{(s)}_{[1,1,1,1]}+I^{(u)}_{[1,1,1,1]}\Big)
\notag
\end{align}
In the small-angle approximation, we sum the tree-level amplitude in eq.\eqref{tre-polsma-2} with the one-loop amplitude in eq.\eqref{loop-th}. 
Keeping terms up to $ \mathcal{O}(\theta^0) $, and applying eq.\eqref{man-aprx}, we obtain 
\begin{align}
i\mca{M}^{\theta^0}_{\mrm{1-loop}}&
{=}-\frac{\kappa^2m^2\omega^2}{\bfq^2}
{-}\frac{15 \kappa^4m^3\omega^2}{512\sqrt{\bfq^2}}
\label{amp-th-0}
\\
&\hspace{-15mm}
{-}\frac{\kappa^4m^2\omega^2}{32\pi^2}\log^2\big(\frac{\bfq^2}{\bar{\mu}^2}\big)
+\frac{\mca{B} \kappa^4m^2\omega^2}{(8\pi)^2}\log\big(\frac{\bfq^2}{\bar{\mu}^2}\big)
\notag
\\
&\hspace{-15mm}
{-}\frac{15\kappa^4m^2\omega^2}{512\pi^2}\log\big(\frac{\bfq^2}{m^2}\big){+}\frac{i\kappa^4m^3\omega^3}{8\pi{}t}\log\big(\frac{\bfq^2}{m^2}\big),
\notag
\end{align}
where $\mca{B}=-\frac{61}{24}$ and we use the simplified version of the scalar box integral in the low energy limit
\begin{equation}\label{boxIntSim}
I^{(s)}_{[1,0,1,1]}+I^{(u)}_{[1,0,1,1]}\simeq-\frac{i}{16\pi^2}\frac{2i\pi}{2tm\omega}\log(-\frac{m^2}{t}).
\end{equation}
The value of $\mathcal{B}$ depends on several ingredients. We construct our perturbation field theory on Minkowskian spacetime, and so do not distinguish between gravitons connecting with external particles and those flowing inside loops. 
At the same time we include the gravi-ghost exchange contributions, diagram (13) of Figure \ref{figOneLoopDiagrams} in our scattering amplitude, as Refs.\cite{Bjerrum-Bohr-2013,Holstein-2016} do.  
Ref. \cite{Bohr-2016} yields $\mathcal{B}=-\frac{161}{120}$, but they construct their gravitational perturbation field theory on curved spacetime, distinguishing between gravitons connecting with external lines and those flowing inside loops. 
Ref. \cite{Bai-2017.03} calculates the same diagrams as ours but uses background field perturbation theory and yields $\mathcal{B}=\frac{113}{120}$. 
Ref. \cite{Chi-2019} considers contributions from the photon but not the gravi-ghost exchange diagrams and yields $\mathcal{B}=\frac{113}{120}$. Refs. \cite{Bai-2017.03} and \cite{Chi-2019} agree since
they perform the same calculation, while ref.\cite{Bohr-2016} is different because it
includes only the two-graviton intermediate states.
The origin of discrepancies between our results relative to these literatures will be analyzed further in \secref{Discuss}.

\section{The deflection angle of photon trajectory}\label{chpt:ang}

In this section, we provide the detailed calculation of the deflection angle of photon trajectories from eq.\eqref{amp-th-0}.

\subsection{Geometric optics approximation}

Since the wavelength $\lambda$ of the photon is much smaller than the horizon scale $2G_N m$ of the black hole, we can describe the propagation of photons with geometric optics. 
According to ref.\cite{Brau-2003}, this is controlled by
\begin{equation}\label{diff-phoprop}
\frac{\rmd}{\rmd s} \left(n\frac{\rmd \mbf{r}}{\rmd s}\right)= \nabla n,
\end{equation}
where $\mathbf{r}$ is the position vector of the photon, $s$ is the affine parameter along its path, and $n$ is the refractive index.  
We choose the coordinate time $t$ as the affine parameter for the photon, so that the above equation becomes:
\begin{equation}\label{diff-proppho-1}
\frac{1}{c^2}\frac{\rmd^2 \mbf{r}}{\rmd t^2}=\frac{1}{n}\nabla n.
\end{equation}
The gravitational field deflecting our photon trajectories is that of the standard Schwarzschild black hole
\begin{align}
\rmd s^2&{=} -h(r) \rmd t^2{+}h^{-1}(r) \rmd r^2{+}r^2(\rmd\theta^2{+}\sin^2\theta \rmd \varphi^2),
\label{sch-metric}
\\
h&= 1-\frac{2G_N m}{r} .
\label{g00}
\end{align}
According to ref.\cite{Alsing-1998}, in the spacetime described by the Schwarzschild metric \eqref{sch-metric}, the refractive index of \eqref{diff-proppho-1}  can be written in the following form,
\begin{align}
n=\sqrt{\frac{1}{h^2}}=&\frac{1}{1-\frac{2G_N m}{r}} \simeq{}1- V_\mathrm{eff}(r) .
\label{ntoV-1}
\end{align}
By a Fourier transformation of \eqref{amp-th-0}, we can obtain the effective potential $\omega\cdot V_{\mrm{eff}}$ of the photon-black hole interaction. 
From eq.\eqref{ntoV-1}, we know that in the region $ r \gg 2G_N m $, $ n $ can be expressed as a function of the effective gravitational potential,
\begin{align}
\omega V_{\mrm{eff}}(r)=&{-}\frac{2 G_N m \omega }{r}{-}\frac{15G_N^2m^2\omega}{4r^2}{+}\frac{2G_N^2\mca{B}~m \omega}{\pi{}r^3}
\notag
\\
&
{-}\frac{G_N^2m\omega}{4\pi{}r^3}{-}\frac{16G_N^2m\omega}{\pi{}r^3}\log\frac{r}{b_0}
{+}\mathcal{O}(\frac{1}{r^4}).
\label{Veff}
\end{align}
where $b_0^{-1}=\mathrm{e}^{1-\gamma_E}\bar{\mu}$ is the infrared cutoff mass. 
Substituting eqs.\eqref{ntoV-1}\&\eqref{Veff} into \eqref{diff-proppho-1}, and performing the integration over $t$, we get
\begin{align}
\frac{1}{c^2} \frac{\rmd \mbf{r}}{\rmd t} = - \frac{1}{\omega}\int\rmd t~\nabla (\omega V_{\mrm{eff}}).
\label{phoMomtChange-1}
\end{align}
\begin{figure}[htbp]
\centering
\includegraphics[totalheight=35mm]{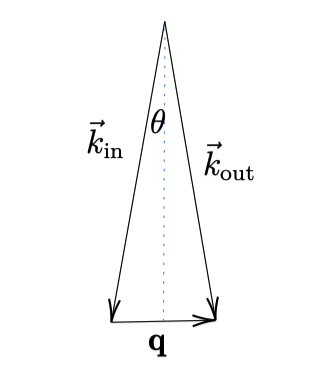}
\includegraphics[totalheight=35mm]{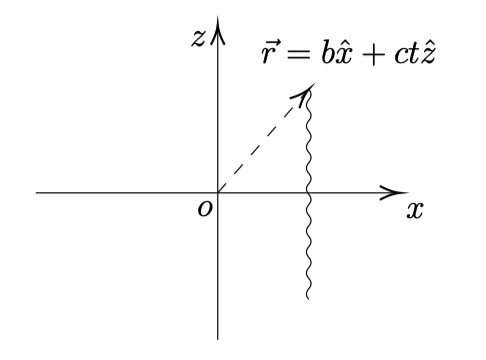}  
\caption{The left panel shows the photon momenta, while the right panel gives the trajectory of photons without external potentials. 
The wavy line in the right panel is the photon trajectory and $b$ is the impact parameter relative to the central black hole.}
\label{fig:sma&pho-tra}
\end{figure}

In the absence of gravitational potentials, the position vector of the photon can be approximately written as $\mathbf{r}=b \hat{\mathbf{x}} + c t \hat{\mathbf{z}}$, as shown in \figref{fig:sma&pho-tra}. 
The deflection angle of the photon’s outgoing momentum $\vec{k}_{\text{out}}$ relative to the incoming momentum $\vec{k}_{\text{in}}$ is very small, as shown in the left panel of \figref{fig:sma&pho-tra}. 
Similarly, the deflection angle $\theta$ of the photon’s position vectors $\mathbf{r}_{\text{in}}$ and $\mathbf{r}_{\text{out}}$ is also small. 
Therefore,
\begin{align}
&\frac{1}{c}\sin\frac{\theta}{2}=-{\int}\rmd t~\nabla V_{\mrm{eff}}(r)
\notag
\\
&\Rightarrow
\theta \simeq{-}{\int}\rmd ct~V'_{\mrm{eff}}(\sqrt{b^2+c^2t^2})\frac{b}{\sqrt{b^2+c^2t^2}}.
\label{phoMomtChange-3}
\end{align}
Let $t=\frac{bu}{c}$, we can easily show that
\begin{align}
\theta &=-b \int_{-\infty}^{+\infty}\rmd u~V'_{\mrm{eff}}(b\sqrt{1+u^2})\frac{1}{\sqrt{1+u^2}}
\label{geo-th-1}
\\
&=\frac{4 G_Nm}{b}{+}\frac{15\pi{}G_N^2m^2}{4b^2}
{+}\frac{G_N^2m\hbar}{\pi{}b^3}\Big(-8 \mca{B}{+}17
\label{geo-th-2}
\\
&\hspace{5mm}
{+}64\log\frac{b}{2b_0}{+}\mathcal{O}(\frac{1}{b^4}).
\notag
\end{align}
This is our photon deflection angle derived from the geometric optics approximation.

\subsection{Eikonal approximation}

Before applying the eikonal approximation, we need to point out a key feature of the transferred momentum $ q^{\mu} $. 
According to the on-shell condition for the external photon, we have
\beq{onsh-pho}
k_{\mrm{i}}\cdot k_{\mrm{i}}=k_{\mrm{o}}\cdot k_{\mrm{o}}=0.
\eeq
If we assume that \( q^{\mu} \) has a non-zero component in the \( z \)-direction, then the outgoing photon momentum can be written as \( k_{\text{o}}^{\mu} = (\omega, q_x, q_y, \omega + q_z) \). 
In this case, taking the square of $k_{\text{o}}^{\mu}$ and applying eq.\eqref{onsh-pho}, we will have
\begin{align}
&0=\omega^2-q_x^2-q_y^2-(\omega+q_z)^2
\label{cha-q-1} 
\\
&\hspace{1.75mm}
=q_x^2+q_y^2+q_z^2 -2\omega \cdot q_z
\notag
\\
&\hspace{-6mm}\Rightarrow\bfq^2=2 \mbf{k}_{\mrm{i}}\cdot \bfq
,
|\bfq|=2\mbf{k}_{\mrm{i}}{\cdot}\hat{\bfq} = 2 \omega \cos\theta.
\label{cha-q-4} 
\end{align}
The low-energy approximation requires $ |\mathbf{q}| \ll \omega $, which implies that $ \cos\theta \ll 1 $ according to eq.\eqref{cha-q-4}. 
Since $ \mathbf{k}_{\mrm{i}} $ only has a component along the $ z $-axis, eq.\eqref{cha-q-4} shows that $ \mathbf{q} $ has almost no component along $ \mathbf{k}_{\mrm{i}} $, the momentum of the ingoing photon. 
In other words, the transferred momentum of the scattering is dominated by the transverse component. 
According to \cite{Bohr-2016, Akhoury-2021.03}, we can write the scattering amplitude in the collisional space as
\begin{equation}\label{amp-b}
\tilde{\mca{M}}(\bfb) = \int \frac{\rmd^2 \bfq}{(2\pi)^2}~\mca{M}(\bfq) e^{-i\bfq\cdot\bfb}.
\end{equation}
Applying the eikonal approximation, this expression can be written in the following form,
\begin{equation} \label{eko-amp}
i\tilde{\mca{M}}(\bfb)\simeq{}2(s{-}m^2)\mathrm{e}^{\big[i\tilde{M}_{\mrm{t}}(\bfb){+}i \tilde{M}_{\mrm{1-loop}}(\bfb)\big]}{-}1,
\end{equation}
where
\begin{align}
\tilde{M}_{\mrm{t}}(\bfb) &=\frac{1}{4m\omega} \int \frac{\rmd^2 \bfq}{(2\pi)^2}~\mca{M}_{\mrm{t}}(\bfq) e^{i\bfq\cdot\bfb},
\label{treamp-b}
\\
\tilde{M}_{\mrm{1-loop}}(\bfb) &=\frac{1}{4m\omega} \int \frac{\rmd^2 \bfq}{(2\pi)^2}~\mca{M}_{\mrm{1-loop}}(\bfq) e^{i\bfq\cdot\bfb} .
\label{loopamp-b}
\end{align}
According to \cite{Akhoury-2021.03}, we can calculate the scattering angle by the stationary phase approximation. 
Considering the calculation of the scattering amplitude in the transferred momentum space
\begin{align}
&\int \rmd^2\bfb~i\mca{M}(\bfb)~\mrm{e}^{-i\bfq\cdot\bfb}
\label{ekophase-2}
\\
&\hspace{-7mm}{=}{\int}\rmd^2\bfb\Big\{2 (s{-}m^2)\mathrm{e}^{\big[i\tilde{M}_{\mrm{t}}(\bfb){+}i\tilde{M}_{\mrm{1-loop}}(\bfb)\big]}{-}1\Big\}\mrm{e}^{-i\bfq\cdot\bfb},
\notag
\end{align}
the stationary phase approximation says that the main contribution of the right-hand integration comes from the region around $\frac{\partial \, \mathrm{phase}}{\partial b} = 0$,
\begin{align}
0&=\frac{\partial}{\partial{}b}\left[{-}\bfq\cdot\bfb{+}\tilde{M}_{\mrm{t}}(\bfb){+}\tilde{M}_{\mrm{1-loop}}(\bfb)\right]
\label{defFromEko-3}
\\
&= 2\omega\sin\frac{\theta}{2}{+}\frac{\partial\tilde{M}_{\mrm{t}}(\bfb)}{\partial b}{+}\frac{\partial\tilde{M}_{\mrm{1-loop}}(\bfb)}{\partial{}b}.
\notag
\end{align}
According to \figref{fig:sma&pho-tra}, we express $q$ here as $ 2\omega \sin\frac{\theta}{2} $.
Performing the series expansion of eqs.\eqref{tre-polsma-2} and \eqref{loop-th} up to $\mca{O}(\theta^0)$, translating into impact parameter space with eqs.\eqref{treamp-b}-\eqref{loopamp-b} and substituting the resultant amplitude into eq.\eqref{defFromEko-3}, we will obtain
\begin{align}
\theta&\simeq-\frac{1}{\omega}\Big[\frac{\partial\tilde{M}^{\theta^0}_{\mrm{t}}(\bfb)}{\partial{}b}+\frac{\partial\tilde{M}^{\theta^0}_{\mrm{1-loop}}(\bfb)}{\partial b}\Big] 
\label{defFromEko-4}
\\
&=\frac{4 G_N m}{b}{+}\frac{15\pi G_N^2 m^2}{4 b^2}{+}\frac{G_N^2 m \hbar}{\pi{}b^3}
\Big(-8{}\mca{B}{+}17
\label{eko-th}
\\
&\hspace{5mm}
{+}64\log\frac{b}{2b_0}\Big){+}\mathcal{O}(\frac{1}{b^4}).
\notag
\end{align}
This is our photon deflection angle derived from the eikonal approximation.

\subsection{Discussion on the result}\label{Discuss}

This section is devoted to comparing our deflection angle expressions in eqs.\eqref{geo-th-2} and \eqref{eko-th} with those quoted in the literature. 
For the classical part, our result can be compared with those derived in classical general relativity (GR) and the Gauss-Bonnet (GB) topological method \cite{Virbhadra-1998,Virbhadra-2000.09},\cite{Gibbons-2008,Heidari-2024}
\begin{align}
\theta_{\mrm{GR}}&\simeq{}\frac{4 G_N m}{b}{+}\frac{4 G_N^2 m^2}{b^2}\Big(\frac{15\pi}{16}{-}1\Big)
\label{resuGR}
\\
&\hspace{5mm}{+}\mca{O}\Big( \frac{G_N^3 m^3}{b^3} \Big),
\notag
\\
\theta_{\mrm{GB}}&\simeq{}\frac{4G_N{}m}{b}{+}\frac{3\pi{}G_N^2{}m^2}{4b^2}.
\label{resuGB}
\end{align}
Obviously, at leading order, our calculation is consistent with results from classical GR and the GB-topological method. 
At next to leading order, our result is nearly twice as large as that from GR and five times as large as that from GB-topological method. 
Nevertheless, the classical part of our result matches with results from references \cite{Bai-2017.03, Brunello-2022, Bohr-2016}, which calculate the deflection angle via the EFT method. 
Results based on EFT contain quantum corrections arising from the quantum feature of gravitational interactions between the photon and the black hole, rather than the quantum modifications of spacetime geometry \cite{Heidari-2024, Zhang-2024}.

However, the quantum part in our results \eqref{geo-th-2} and \eqref{eko-th} does not match those of references \cite{Bai-2017.03, Brunello-2022, Bohr-2016} exactly. 
The differences and their origins are as follows.

\begin{enumerate}

\item The coefficient structure of the non-$\log$ part of our results $(-8\mca{B} + \mathcal{C})$, is the same as those in \cite{Bohr-2016,Bai-2017.03,Brunello-2022,Chi-2019}. 
However, just as we pointed out under eq\eqref{boxIntSim}, different works yield different values for the parameter $\mca{B}$. 
Our calculation yields $\mca{B}=-\frac{61}{24}$; ref.\cite{Bai-2017.03} yields $\mca{B} =-\frac{39}{8}$; ref.\cite{Brunello-2022} yields $\mca{B}=\frac{3}{40}$ for the scalar photon scatterings; ref. \cite{Bohr-2016} yields $\mca{B}=\frac{3}{40}$, $-\frac{31}{30}$ and $-\frac{161}{120}$ for scalar, spinor and vector photon scatterings respectively. 
Although the equivalence principle requires that these parameters be universal, ref.\cite{Bohr-2016} argues that the truncated series expansion of an EFT has no reason to capture the full geometric features of GR. 
One reason for these differences is because different authors have considered different particle exchanges in the loop diagrams, just as ref.\cite{Chi-2019} pointed out explicitly. 

\item The most important reason for the difference in $\mathcal{B}$ between our results and those of \cite{Bai-2017.03,Brunello-2022,Bohr-2016,Chi-2019} originates from the differences between the Feynman rules we adopted for the three-graviton vertex and those adopted by these works. 
Our rules are based on a simple perturbation treatments of the quantum gravitational field \cite{Prinz-2021,Latosh-2022,Latosh-2023.11}, which makes no delineation between gravitons carrying only external momentums and those carrying loop momentum. 
Meanwhile, the calculation of \cite{Bai-2017.03,Brunello-2022,Bohr-2016,Chi-2019} uses the Feynman rules \cite{Donoghue-1994.05,Donoghue-1994.09,Bohr-2015} of perturbative field theory on a non-trivial background. 
In this background field perturbation method, the Feynman rules distinguish two types of gravitons; those carrying only external momentum and those flowing loop momentum. 
In the simple perturbation method for quantum gravitational field, the three graviton vertex related Lagrangian \cite{MG-1984} can be written as

\begin{widetext}
\begin{align}
\mathcal{L}_{h}&=\sqrt{-g}
\frac{1}{2}g^{\mu\nu}g^{\phi\psi}g^{\sigma\rho}\big(
h_{\phi\nu,\sigma}h_{\psi\mu,\rho}
{-}\frac{1}{2}h_{\sigma\rho,\mu}h_{\phi\psi,\nu}
{+}2h_{\mu\sigma,\rho}h_{\nu\phi,\psi}
{-}2h_{\nu\sigma,\phi}h_{\mu\psi,\rho}
\big)
\\
&=\big[\big(
\frac{1}{2}\eta^{\sigma\rho\mu\psi\phi\nu}k{\cdot}p
-\eta^{\mu\sigma\psi\rho\phi\nu}k{\cdot}p
-\eta^{\mu\psi\phi\sigma\nu\rho}k{\cdot}p
-\eta^{\mu\psi\phi\nu}k^\sigma p^\rho
\big)
\\
&\hspace{5mm}-\frac{1}{2}\big(
\frac{1}{2}\eta^{\mu\nu\phi\psi\sigma\rho}k{\cdot}q
-\eta^{\phi\psi\sigma\rho}k^\nu q^\mu
-\eta^{\phi\mu\psi\nu\sigma\rho}k{\cdot}q
-\eta^{\phi\psi\sigma\mu\rho\nu}k{\cdot}q
\big)\nonumber
\\
&\hspace{5mm}+2\big(
\frac{1}{2}\eta^{\mu\nu\rho\psi}k^{\phi}q^{\sigma}
-\eta^{\rho\mu\psi\nu}k^{\phi}q^{\sigma}
-\eta^{\rho\psi\phi\mu}k^\nu q^{\sigma}
-\eta^{\rho\psi\sigma\nu}k^{\phi } q^\mu
\big)\nonumber
\\
&\hspace{5mm}-2\big(
\frac{1}{2}\eta^{\sigma\rho\phi\nu}p^{\psi}k^{\mu}
-\eta^{\phi\sigma\nu\rho}p^{\psi}k^{\mu}
-\eta^{\phi\nu\rho\psi}p^{\sigma}k^{\mu}
-\eta^{\phi\nu\mu\sigma}p^{\psi}k^{\rho}
\big)
\big]h^k_{\phi\psi}h^p_{\mu\nu}h^q_{\sigma\rho}
\nonumber
\\
&\equiv T^{\phi\psi\mu\nu\sigma\rho}_{kpq}h^k_{\phi\psi}h^p_{\mu\nu}h^q_{\sigma\rho},
\label{vertxhhh}
\end{align}
\end{widetext}
where $\eta^{\sigma\rho\mu\nu\phi\psi}\equiv\eta^{\sigma\rho}\eta^{\mu\nu}\eta^{\phi\psi}$, $\eta^{\sigma\rho\mu\nu}\equiv\eta^{\sigma\rho}\eta^{\mu\nu}$. 
According to this Lagrangian, the three graviton vertex can be derived as follows
\begin{equation}
V^{\phi\psi,\mu\nu,\sigma\rho}_{hhh}=\frac{\kappa}{2}\sum_{k\phi\psi,p\mu\nu,q\sigma\rho}^\mathrm{permutation}T^{(\phi\psi)(\mu\nu)(\sigma\rho)}_{kpq}
\end{equation}
where the bracketed indices $(\phi\psi)$, $(\mu\nu)$ $(\sigma\rho)$ are symmetrized.
We calculated the diagrams (4)-(8) in Figure \ref{figOneLoopDiagrams} with both Feynman rules for the three graviton vertex; Bohr's and ours. 
After a series expansion to $\mathcal{O}(t^0,\omega^2)$ the results have the form
\begin{equation}
\mathcal{M}_\mathrm{Bohr}^{(4)-(8)}\propto30m^2 I_{[1,0,1,1]})-43 I_{[1,0,1,0]})
\label{bohr4-8}
\end{equation}
\begin{equation}
\mathcal{M}_\mathrm{our}^{(4)-(8)}\propto30m^2 I_{[1,0,1,1]}-373 I_{[1,0,1,0]}.
\label{our4-8}
\end{equation}
Evidently, different Feynman rules do cause the difference in the coefficient of the $\log\frac{\bfq^2}{\bar{\mu}^2}$ term.

\item The coefficient of the $\log \frac{b}{2b_0}$ term: the expression of \cite{Brunello-2022} based on the geometric optics approximation differs by a sign from that based on the eikonal approximation, but coincides with those of \cite{Bohr-2016}. 
Our expressions based on the two methods show no sign difference, but differ from those of \cite{Bohr-2016} by a sign. 
This is because we used the following Fourier transformation formula
\begin{align}
\int\frac{\rmd^2\bfq}{(2\pi)^2}~\mrm{e}^{-i\bfq\cdot\mbf{r}}~\log^2 \bfq^2&{=}\frac{4}{\pi{}r^2}{+}\frac{4}{\pi{}r^2} \log\frac{r}{2r_0},
\label{equ:ourF2}
\end{align}
where $r_0 = \mrm{e}^{1-\gamma_E}$, while the transformation formula of \cite{Brunello-2022, Bohr-2016} reads as follows
\begin{align}
\int \frac{\rmd^2 \bfq}{(2\pi)^2}~\mrm{e}^{-i \bfq\cdot\mbf{r}}~\log^2 \bfq^2{=}\frac{4}{\pi r^2} \log\frac{2}{r}.
\label{equ:atiF2}
\end{align}
By \eqref{equ:ourF2} the contribution of $\log^2 \frac{\mathbf{q}^2}{\bar{\mu}^2}$ to the long-range potential $V_{\mathrm{eff}}$ is attractive. 
Meanwhile, from \eqref{equ:atiF2} this contribution will be repulsive.
\end{enumerate}
The background field method has the function of capturing non-perturbative effects \cite{Itzykson-1980,Gonoskov-2021,Copinger-2024}, here they refer to geodesic motion in the background spacetime, while the simple perturbation method carries no such effects, the deflection angle discrepancies between the calculations of \cite{Bai-2017.03,Brunello-2022,Bohr-2016,Chi-2019} and ours is to be expected. 
Since scattering happens in the weak field region of the black hole spacetime, the simple perturbation theory of the quantum gravitational field yields almost equally accurate predictions for the deflection angle as the background field theory on Schwarzschild spacetime.  
Our calculation illustrates that their discrepancies manifest at most from the $\mathcal{O}[b^{-3}]$ terms.

\section{Conclusion}\label{chpt:con}

Via the EFT method and in the weak-field limit, we calculated the one-loop amplitude of an electromagnetic field scattering from a heavy scalar field through gravitational interactions. 
The relevant Feynman diagrams are presented in \figref{figOneLoopDiagrams}. 
To simplify the computation, we employed IBP reduction techniques, which reduced the relevant Feynman integrations to four master integrals, significantly reducing the complexity of the loop integral evaluations. 
Focusing on small deflection angles, we isolated the contribution from long-range gravitational interactions by retaining terms in the scattering amplitude up to $\mathcal{O}(t^0, \omega^3)$. 
Utilizing geometric optics and the eikonal approximation, we derived the photon deflection angle caused by massive celestial objects.
 
Unlike traditional QFT approach, we only used the non-analytic terms in $t$ of the scattering amplitude to calculate the deflection angle $\theta$. 
Our calculation incorporates contributions from the graviton-ghost fields to the scattering amplitude. 
The resultant expressions for the deflection angle are compared with those reported in the existing literature. 
Discrepancies are explicitly identified, and their origins were tracked to the difference between the three-graviton vertex Feynman rules adopted in the literature \cite{Donoghue-1994.09,Bohr-2016} and in our calculations \cite{Prinz-2021,Latosh-2022,Latosh-2023.11}. 
Both rules are physical, however the former is based on the background field method, which distinguishes gravitons carrying only external momentum and gravitons carrying loop-momentum, whilst the latter is based on conventional EFT without background field effects. 
Since the scattering happens in the weak field regime, both calculations yield equally accurate predictions for the deflection angle of photons passing by the black hole. 
The difference arises at most from the $\mathcal{O}[b^{-3}]$ terms.

Our work serves as a supplement to the calculation of the deflection angles of light in weak gravitational fields. 
The method employed here is limited to the low-energy regime ($b \gg 2G_N m, t \ll m^2,\omega^2 \ll m^2$) and not applicable to scenarios involving strong gravitational fields. 
For such cases, alternative approaches, such as those developed by V. Bozza and N. Tsukamoto, should be utilized. 
Additionally, we acknowledge that the EFT method may not be the optimal choice for obtaining higher-order correction terms. 
This limitation arises because, as the perturbation order increases, the complexity of the calculation grows significantly. 
Specifically, the number and variety of graviton-matter field interactions and graviton-self-interaction vertices increase dramatically. 
Notably, the appearance of four-graviton self-coupling vertices will introduce expressions of considerable complexity, which makes the EFT approach less practical for such computations.

\acknowledgments

This work is supported by NSFC grant no.11875082. 
We thank Dr. Ikram Ben and Dr. Nick Houston for assisting with language-related revisions.

\appendix

\begin{widetext}
\section{Feynman Rules}
Our Feynman rules are derived from actions \eqref{phi-act}-\eqref{A-act}.  For reproductivity of our computation and for referring convenience of interested readers, we list them out in the following.
Firstly, propagators of the heavy scalar, the photon, the gravi-ghost and the graviton are as follows,
\begin{eqnarray}
\raisebox{-0.3\height}{\includegraphics[totalheight=13mm]{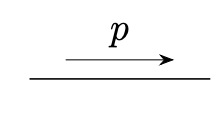}}
&&=-\frac{i}{p^2-m^2}
\\
\raisebox{-0.3\height}{\includegraphics[totalheight=13.5mm]{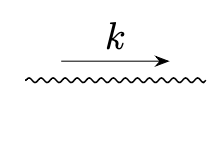}}
&&=-\frac{i\eta_{\mu\nu}}{k^2}
\\
\raisebox{-0.3\height}{\includegraphics[totalheight=10mm]{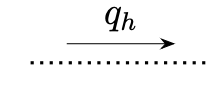}}
&&=\frac{i\eta_{\mu\nu}}{k^2}
\\
\raisebox{-0.3\height}{\includegraphics[totalheight=11.5mm]{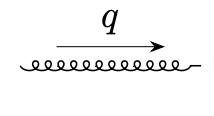}}
&&=\frac{1}{2}
\frac{\eta_{\mu\rho}\eta_{\nu\sigma}+\eta_{\mu\sigma}\eta_{\nu\rho}-\eta_{\mu\nu}\eta_{\rho\sigma}}{q^2}
\end{eqnarray}
Then, all the relevant vertices needed are as follows:
\begin{eqnarray}
\raisebox{-0.5\height}{\includegraphics[totalheight=17mm]{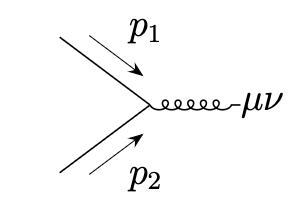}}
=\frac{i}{2}\kappa{}(p_1^{\mu}p_2^{\nu}
+p_1^{\nu}p_2^{\mu}
-p_1{\cdot}p_2g^{\mu\nu}
-m^2 g^{\mu \nu }
),
\label{fr:verPPH}
\end{eqnarray}
\begin{eqnarray}
\raisebox{-0.5\height}{\includegraphics[totalheight=17mm]{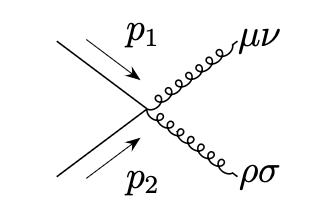}}
=\frac{i}{2}\kappa^2(v_{p_1p_2}^{\mu\nu\rho\sigma}+v_{p_1p_2}^{\rho\sigma\mu\nu}
+v_{p_2p_1}^{\mu\nu\rho\sigma}+v_{p_2p_1}^{\rho\sigma\mu\nu}),
\label{fr:verPPHH}
\end{eqnarray}

\begin{eqnarray}
v_{p_1p_2}^{\mu\nu\rho\sigma}\equiv\Big(
{-}\frac{1}{4}m^2\eta^{\mu\nu\rho\sigma}
{+}\frac{1}{4}m^2\eta^{\mu\rho\nu\sigma}
{+}\frac{1}{4}m^2\eta^{\nu\rho\mu\sigma}
{+}\frac{1}{2}p_1^{\mu}p_2^{\nu}\eta^{\rho\sigma}
\label{LagPPHH}
\\
{-}\frac{1}{2}p_1^{\mu}p_2^{\rho}\eta^{\nu\sigma}
{-}\frac{1}{2}p_1^{\mu}p_2^{\sigma}\eta^{\nu\rho}
{+}\frac{1}{2}p_2^{\mu}p_1^{\nu}\eta^{\rho\sigma}
{-}\frac{1}{2}p_1^{\nu}p_2^{\rho}\eta^{\mu\sigma}
\nonumber
\\
{-}\frac{1}{2}p_1^{\nu}p_2^{\sigma}\eta^{\mu\rho}
{-}\frac{1}{4}p_1{\cdot}p_2\eta^{\mu\nu\rho\sigma}
{+}\frac{1}{4}p_1{\cdot}p_2\eta^{\mu\rho\nu\sigma}
{+}\frac{1}{4}p_1{\cdot}p_2\eta^{\nu\rho\mu\sigma}\Big),
\nonumber
\end{eqnarray}

\begin{eqnarray}
\raisebox{-0.5\height}{\includegraphics[totalheight=17mm]{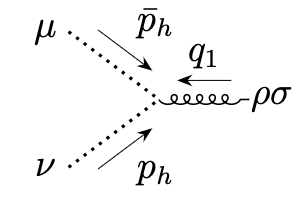}}
=\frac{i\kappa}{2}\Big(
{-}\bar{p}_h^{\sigma}\eta^{\mu\nu}p_h^{\rho}
{-}\bar{p}_h^{\rho}\eta^{\mu\nu}p_h^{\sigma}
{+}\bar{p}_h^{\alpha}p_{h\alpha}\eta^{\mu\nu\rho\sigma}
{-}\bar{p}_h^{\alpha}p_{h\alpha}\eta^{\mu\rho\nu\sigma}
\\
{-}\bar{p}_h^{\alpha}p_{h\alpha}\eta^{\nu\rho\mu\sigma}
{-}\frac{1}{2}q_1^{\mu}\bar{p}_h^{\rho}\eta^{\nu\sigma}
{-}\frac{1}{2}q_1^{\mu}\bar{p}_h^{\sigma}\eta^{\nu\rho}
{+}\frac{1}{2}q_1^{\nu}\bar{p}_h^{\rho}\eta^{\mu\sigma}
\notag
\\
{+}\frac{1}{2}q_1^{\nu}\bar{p}_h^{\sigma}\eta^{\mu\rho}
{-}\frac{1}{2}\bar{p}_h{\cdot}q_1\eta^{\mu\rho\nu\sigma}
{-}\frac{1}{2}\bar{p}_h{\cdot}q_1\eta^{\nu\rho\mu\sigma}
{+}\frac{1}{2}q_1^{\mu}\eta^{\nu\sigma}p_h^{\rho}
\notag
\\
{+}\frac{1}{2}q_1^{\mu}\eta^{\nu\rho}p_h^{\sigma}
{-}\frac{1}{2}q_1^{\nu}\eta^{\mu\sigma}p_h^{\rho}
{-}\frac{1}{2}q_1^{\nu}\eta^{\mu\rho}p_h^{\sigma}
{-}\frac{1}{2}p_h{\cdot}q_1\eta^{\mu\rho\nu\sigma}
\notag
\\
{-}\frac{1}{2}p_h{\cdot}q_1\eta^{\nu\rho\mu\sigma}
{+}\eta^{\rho\sigma}q_1^{\mu}q_1^{\nu}
{-}\frac{1}{2}\eta^{\nu\sigma}q_1^{\mu}q_1^{\rho}
{-}\frac{1}{2}\eta^{\mu\sigma}q_1^{\nu}q_1^{\rho}
\notag
\\
{-}\frac{1}{2}\eta^{\nu\rho}q_1^{\mu}q_1^{\sigma}
{-}\frac{1}{2}\eta^{\mu\rho}q_1^{\nu}q_1^{\sigma}
{+}\frac{1}{2}q_1{\cdot}q_1\eta^{\mu\rho\nu\sigma}
{+}\frac{1}{2}q_1{\cdot}q_1\eta^{\nu\rho\mu\sigma}
\Big),
\notag
\end{eqnarray}

\begin{eqnarray}
\raisebox{-0.5\height}{\includegraphics[totalheight=19mm]{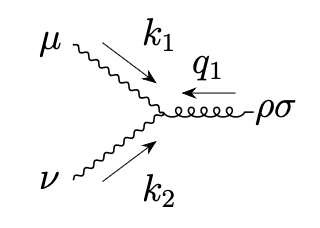}}
=\frac{i}{2}\kappa~\big(
{-}\frac{1}{2}k_2^{\mu}k_1^{\nu}\eta^{\rho\sigma}
{+}k_1^{\nu}k_2^{\rho}\eta^{\mu\sigma}
{-}\frac{1}{2}k_2^{\rho}k_1^{\sigma}\eta^{\mu\nu}
{+}k_1^{\nu}k_2^{\sigma}\eta^{\mu\rho}
\label{fr:verAAH}
\\
{-}\frac{1}{2}k_1^{\rho}k_2^{\sigma}\eta^{\mu\nu}
{+}\frac{1}{2}k_1{\cdot}k_2\eta^{\mu\nu\rho\sigma}
{-}\frac{1}{2}k_1{\cdot}k_2\eta^{\mu\rho\nu\sigma}
{-}\frac{1}{2}k_1{\cdot}k_2\eta^{\nu\rho\mu\sigma}
\big)+\big(\mu\&k_1 \leftrightarrow \nu\&k_2\big),
\notag
\end{eqnarray}

\begin{eqnarray}
\raisebox{-0.5\height}{\includegraphics[totalheight=17mm]{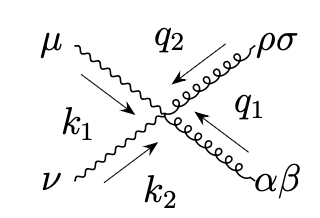}}
&&=
\frac{i}{2}\kappa^2~\big(
v_{k_1k_2}^{\mu\nu\rho\sigma\alpha\beta}
+v_{k_2k_1}^{\nu\mu\rho\sigma\alpha\beta}
+v_{k_1k_2}^{\mu\nu\alpha\beta\rho\sigma}
+v_{k_2k_1}^{\nu\mu\alpha\beta\rho\sigma}
\big),
\end{eqnarray}

\begin{eqnarray}
v_{k_1k_2}^{\mu\nu\rho\sigma\alpha\beta}\equiv\Big(
\frac{1}{8}k_1{\cdot}k_2\eta^{\rho\sigma}\eta^{\alpha\beta\mu\nu}
{-}\frac{1}{4}k_1{\cdot}k_2\eta^{\nu\sigma}\eta^{\alpha\beta\mu\rho}
{-}\frac{1}{4}k_1{\cdot}k_2\eta^{\mu\sigma}\eta^{\alpha\beta\nu\rho}
{+}\frac{1}{4}k_1{\cdot}k_2\eta^{\nu\sigma}\eta^{\alpha\mu\beta\rho}
\\
{+}\frac{1}{4}k_1{\cdot}k_2\eta^{\beta\sigma}\eta^{\alpha\mu\nu\rho}
{+}\frac{1}{4}k_1{\cdot}k_2\eta^{\nu\sigma}\eta^{\beta\mu\alpha\rho}
{+}\frac{1}{4}k_1{\cdot}k_2\eta^{\alpha\sigma}\eta^{\beta\mu\nu\rho}
{-}\frac{1}{8}k_1{\cdot}k_2\eta^{\beta\sigma}\eta^{\mu\nu\alpha\rho}
\notag
\\
{-}\frac{1}{8}k_1{\cdot}k_2\eta^{\alpha\sigma}\eta^{\mu\nu\beta\rho}
{-}\frac{1}{4}k_1^{\alpha}k_2^{\sigma}\eta^{\beta\nu\mu\rho}
{-}\frac{1}{4}k_1^{\alpha}k_2^{\rho}\eta^{\beta\nu\mu\sigma}
{+}\frac{1}{4}k_1^{\alpha}k_2^{\beta}\eta^{\mu\rho\nu\sigma}
\notag
\\
{+}\frac{1}{4}k_1^{\alpha}k_2^{\beta}\eta^{\nu\rho\mu\sigma}
{-}\frac{1}{4}k_1^{\beta}k_2^{\sigma}\eta^{\alpha\nu\mu\rho}
{-}\frac{1}{4}k_1^{\beta}k_2^{\rho}\eta^{\alpha\nu\mu\sigma}
{+}\frac{1}{4}k_2^{\alpha}k_1^{\beta}\eta^{\mu\rho\nu\sigma}
\notag
\\
{+}\frac{1}{4}k_2^{\alpha}k_1^{\beta}\eta^{\nu\rho\mu\sigma}
{+}\frac{1}{2}k_1^{\nu}k_2^{\sigma}\eta^{\alpha\beta\mu\rho}
{+}\frac{1}{2}k_1^{\nu}k_2^{\rho}\eta^{\alpha\beta\mu\sigma}
{-}\frac{1}{8}k_2^{\mu}k_1^{\nu}\eta^{\alpha\beta\rho\sigma}
\notag
\\
{+}\frac{1}{8}k_2^{\mu}k_1^{\nu}\eta^{\alpha\rho\beta\sigma}
{-}\frac{1}{2}k_2^{\beta}k_1^{\nu}\eta^{\alpha\rho\mu\sigma}
{+}\frac{1}{8}k_2^{\mu}k_1^{\nu}\eta^{\beta\rho\alpha\sigma}
{-}\frac{1}{2}k_2^{\alpha}k_1^{\nu}\eta^{\beta\rho\mu\sigma}
\notag
\\
{-}\frac{1}{2}k_2^{\beta}k_1^{\nu}\eta^{\mu\rho\alpha\sigma}
{-}\frac{1}{2}k_2^{\alpha}k_1^{\nu}\eta^{\mu\rho\beta\sigma}
{-}\frac{1}{4}k_1^{\rho}k_2^{\sigma}\eta^{\alpha\beta\mu\nu}
{+}\frac{1}{4}k_2^{\beta}k_1^{\rho}\eta^{\mu\nu\alpha\sigma}
\notag
\\
{+}\frac{1}{4}k_2^{\alpha}k_1^{\rho}\eta^{\mu\nu\beta\sigma}
{-}\frac{1}{4}k_2^{\rho}k_1^{\sigma}\eta^{\alpha\beta\mu\nu}
{+}\frac{1}{4}k_2^{\beta}k_1^{\sigma}\eta^{\mu\nu\alpha\rho}
{+}\frac{1}{4}k_2^{\alpha}k_1^{\sigma}\eta^{\mu\nu\beta\rho}\Big),
\notag
\end{eqnarray}
\begin{eqnarray}
\raisebox{-0.5\height}{\includegraphics[totalheight=17mm]{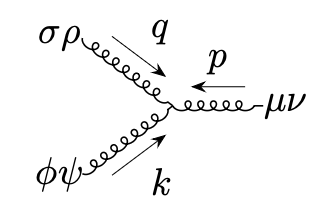}}
=\sum^\mathrm{permutation}_{(k\phi\psi)(p\mu\nu)(q\sigma\rho)}
\frac{i}{2}\kappa~v^{\phi\psi\mu\nu\sigma\rho}_{hhh}.
\end{eqnarray}
where $v^{\phi\psi\mu\nu\sigma\rho}_{hhh}$ has been given in eqs.\eqref{vertxhhh}.
\end{widetext}

\section{Master Intgrals}

Expressions of the master integrals involved in the Feynman diagrams of \figref{figOneLoopDiagrams} in terms of the Mandelstam variables are provided below for readers who are interested in reproducing or checking our calculations,
\begin{align}
I_{[1,0,1,0]}&\equiv\ddq\frac{1}{D_1 D_3}
\label{mi-1}\\
&\simeq\frac{i}{16 \pi^2}\Big[\log\frac{-\mu^2}{t}{+}\frac{\epsilon}{2}\log^2\frac{-\mu^2}{t}\Big],
\notag
\end{align}
\begin{align}
I_{[1,0,1,1]}&{\equiv}\ddq\frac{1}{D_1 D_3 D_4}
\notag
\\
&\simeq\frac{i}{32 \pi^2 m^2}\Big[{-}\frac{\pi^2 m}{\sqrt{{-}t}}+\log\frac{{-}m^2}{t}
\Big],
\label{mi-2}
\end{align}
\begin{align}
&I^{(s)}_{[1,1,1,1]}\equiv\ddq\frac{1}{D_1 D_2 D_3 D_4}
\label{mi-3}
\\
&\hspace{-1mm}
\simeq\frac{i}{16\pi^2}\frac{1}{t(m^2{-}s)}\Big(\frac{\mu^2}{m^2}\Big)^{\epsilon}\Big\{{-}\frac{2}{\epsilon^2}{-}\frac{1}{\epsilon}\Big[2\log\frac{m^2}{m^2{-}s}{+}
\notag
\\
&\hspace{-1mm}
\log({-}\frac{m^2}{t})\Big]-\log(\frac{m^2}{m^2{-}s})\log({-}\frac{m^2}{t})+\frac{\pi^2}{2}\Big\},
\notag
\end{align}
\begin{align}
&I^{(u)}_{[1,1,1,1]}\equiv\ddq\frac{1}{D_1 D_2 D_3 D_5} 
\label{itegralU1111}
\\
&\hspace{-1mm}
\simeq\frac{i}{16\pi^2}\frac{1}{t(m^2{-}u)}\Big(\frac{\mu^2}{m^2}\Big)^{\epsilon}\Big\{{-}\frac{2}{\epsilon^2}{-}\frac{1}{\epsilon}\Big[2\log\frac{m^2}{m^2{-}u}{+}
\notag
\\
&\hspace{-1mm}
\log({-}\frac{m^2}{t})\Big]-\log(\frac{m^2}{m^2{-}u})\log({-}\frac{m^2}{t})+\frac{\pi^2}{2}\Big\}.
\notag
\end{align}

\newpage

\addcontentsline{toc}{section}{\bf Reference}
\bibliographystyle{unsrt}

\bibliography{deflect-ref}

@article{Holstein-2016,
	archiveprefix = {arXiv},
	author = {Holstein, Barry R.},
	doi = {10.1088/0954-3899/44/1/01LT01},
	eprint = {1609.00714},
	journal = {J. Phys. G},
	number = {1},
	pages = {01LT01},
	primaryclass = {hep-ph},
	title = {{Analytical On-shell Calculation of Low Energy Higher Order Scattering}},
	volume = {44},
	year = {2017}}

@article{Bjerrum-Bohr-2013,
	archiveprefix = {arXiv},
	author = {Bjerrum-Bohr, N. E. J. and Donoghue, John F. and Vanhove, Pierre},
	doi = {10.1007/JHEP02(2014)111},
	eprint = {1309.0804},
	journal = {JHEP},
	pages = {111},
	primaryclass = {hep-th},
	reportnumber = {IHES-P-13-23, IPHT-T13-019},
	title = {{On-shell Techniques and Universal Results in Quantum Gravity}},
	volume = {02},
	year = {2014}}

@book{Itzykson-1980,
	address = {New York},
	author = {Itzykson, C. and Zuber, J. B.},
	isbn = {978-0-486-44568-7},
	publisher = {McGraw-Hill},
	series = {International Series In Pure and Applied Physics},
	title = {{Quantum Field Theory}},
	year = {1980}}

@article{Copinger-2024,
	archiveprefix = {arXiv},
	author = {Copinger, Patrick and Edwards, James P. and Ilderton, Anton and Rajeev, Karthik},
	doi = {10.1103/PhysRevD.111.036009},
	eprint = {2411.06203},
	journal = {Phys. Rev. D},
	number = {3},
	pages = {036009},
	primaryclass = {hep-ph},
	title = {{Pair creation, backreaction, and resummation in strong fields}},
	volume = {111},
	year = {2025}}

@article{Gonoskov-2021,
	archiveprefix = {arXiv},
	author = {Gonoskov, A. and Blackburn, T. G. and Marklund, M. and Bulanov, S. S.},
	doi = {10.1103/RevModPhys.94.045001},
	eprint = {2107.02161},
	journal = {Rev. Mod. Phys.},
	number = {4},
	pages = {045001},
	primaryclass = {physics.plasm-ph},
	title = {{Charged particle motion and radiation in strong electromagnetic fields}},
	volume = {94},
	year = {2022}}

@article{Prinz-2021,
	author = {Prinz, David},
	doi = {10.1088/1361-6382/ac1cc9},
	issn = {1361-6382},
	journal = {Classical and Quantum Gravity},
	month = oct,
	number = {21},
	pages = {215003},
	publisher = {IOP Publishing},
	title = {Gravity-Matter Feynman Rules for any Valence},
	url = {http://dx.doi.org/10.1088/1361-6382/ac1cc9},
	volume = {38},
	year = {2021}}

@book{MG-1984,
	author = {Dubrovin, B. A. and Fomenko, A. T. and Novikov, S. P.},
	doi = {10.1007/978-1-4684-9946-9},
	publisher = {Springer},
	title = {Modern Geometry --- Methods and Applications},
	url = {https://app.dimensions.ai/details/publication/pub.1006644859},
	year = {1992}}

@article{Bohr-2015,
	author = {Bjerrum-Bohr, N. E. J. and Holstein, Barry R. and Plant\'e, Ludovic and Vanhove, Pierre},
	doi = {10.1103/PhysRevD.91.064008},
	issue = {6},
	journal = {Phys. Rev. D},
	month = {Mar},
	numpages = {18},
	pages = {064008},
	publisher = {American Physical Society},
	title = {Graviton-photon scattering},
	url = {https://link.aps.org/doi/10.1103/PhysRevD.91.064008},
	volume = {91},
	year = {2015}}

@article{Chi-2019,
	author = {Chi, Huan-Hang},
	doi = {10.1103/PhysRevD.99.126008},
	issue = {12},
	journal = {Phys. Rev. D},
	month = {Jun},
	numpages = {7},
	pages = {126008},
	publisher = {American Physical Society},
	title = {Graviton bending in quantum gravity from one-loop amplitudes},
	url = {https://link.aps.org/doi/10.1103/PhysRevD.99.126008},
	volume = {99},
	year = {2019}}

@misc{Zhang-2024,
	archiveprefix = {arXiv},
	author = {Cong Zhang and Jerzy Lewandowski and Yongge Ma and Jinsong Yang},
	eprint = {2407.10168},
	primaryclass = {gr-qc},
	title = {Black Holes and Covariance in Effective Quantum Gravity},
	url = {https://arxiv.org/abs/2407.10168},
	year = {2024}}

@misc{Virbhadra-1998,
	archiveprefix = {arXiv},
	author = {K. S. Virbhadra and D. Narasimha and S. M. Chitre},
	eprint = {astro-ph/9801174},
	primaryclass = {astro-ph},
	title = {Role of the scalar field in gravitational lensing},
	url = {https://arxiv.org/abs/astro-ph/9801174},
	year = {1998}}

@article{EHT-2022,
	author = {Event Horizon Telescope Collaboration et al},
	doi = {10.3847/2041-8213/ac6674},
	issn = {2041-8213},
	journal = {The Astrophysical Journal Letters},
	month = may,
	number = {2},
	pages = {L12},
	publisher = {American Astronomical Society},
	title = {First Sagittarius A* Event Horizon Telescope Results. I. The Shadow of the Supermassive Black Hole in the Center of the Milky Way},
	url = {http://dx.doi.org/10.3847/2041-8213/ac6674},
	volume = {930},
	year = {2022}}

@article{Donoghue-1999.06,
	author = {Donoghue, John F. and Torma, Tibor},
	doi = {10.1103/physrevd.60.024003},
	issn = {1089-4918},
	journal = {Physical Review D},
	month = jun,
	number = {2},
	publisher = {American Physical Society (APS)},
	title = {Infrared behavior of graviton-graviton scattering},
	url = {http://dx.doi.org/10.1103/PhysRevD.60.024003},
	volume = {60},
	year = {1999}}

@article{Donoghue-1994.05,
	author = {Donoghue, John F.},
	doi = {10.1103/physrevlett.72.2996},
	issn = {0031-9007},
	journal = {Physical Review Letters},
	month = may,
	number = {19},
	pages = {2996--2999},
	publisher = {American Physical Society (APS)},
	title = {Leading quantum correction to the Newtonian potential},
	url = {http://dx.doi.org/10.1103/PhysRevLett.72.2996},
	volume = {72},
	year = {1994}}

@article{Garcia-2008.10-1,
	author = {Mart{\'\i}n-Garc{\'\i}a, Jos{\'e} M.},
	doi = {10.1016/j.cpc.2008.05.009},
	issn = {0010-4655},
	journal = {Computer Physics Communications},
	month = oct,
	number = {8},
	pages = {597--603},
	publisher = {Elsevier BV},
	title = {xPerm: fast index canonicalization for tensor computer algebra},
	url = {http://dx.doi.org/10.1016/j.cpc.2008.05.009},
	volume = {179},
	year = {2008}}

@article{Garcia-2008.10-2,
	author = {Mart{\'\i}n-Garc{\'\i}a, J.M. and Yllanes, D. and Portugal, R.},
	doi = {10.1016/j.cpc.2008.04.018},
	issn = {0010-4655},
	journal = {Computer Physics Communications},
	month = oct,
	number = {8},
	pages = {586--590},
	publisher = {Elsevier BV},
	title = {The Invar tensor package: Differential invariants of Riemann},
	url = {http://dx.doi.org/10.1016/j.cpc.2008.04.018},
	volume = {179},
	year = {2008}}

@article{Garcia-2007.10,
	author = {Mart{\'\i}n-Garc{\'\i}a, J.M. and Portugal, R. and Manssur, L.R.U.},
	doi = {10.1016/j.cpc.2007.05.015},
	issn = {0010-4655},
	journal = {Computer Physics Communications},
	month = oct,
	number = {8},
	pages = {640--648},
	publisher = {Elsevier BV},
	title = {The Invar tensor package},
	url = {http://dx.doi.org/10.1016/j.cpc.2007.05.015},
	volume = {177},
	year = {2007}}

@article{Latosh-2023.11,
	author = {Latosh, B.},
	doi = {10.1016/j.cpc.2023.108871},
	issn = {0010-4655},
	journal = {Computer Physics Communications},
	month = nov,
	pages = {108871},
	publisher = {Elsevier BV},
	title = {FeynGrav 2.0},
	url = {http://dx.doi.org/10.1016/j.cpc.2023.108871},
	volume = {292},
	year = {2023}}

@article{Latosh-2022,
	author = {Latosh, B},
	doi = {10.1088/1361-6382/ac7e15},
	issn = {1361-6382},
	journal = {Classical and Quantum Gravity},
	month = jul,
	number = {16},
	pages = {165006},
	publisher = {IOP Publishing},
	title = {FeynGrav: FeynCalc extension for gravity amplitudes},
	url = {http://dx.doi.org/10.1088/1361-6382/ac7e15},
	volume = {39},
	year = {2022}}

@article{Mertig-1991,
	author = {R. Mertig and M. B{\"o}hm and A. Denner},
	doi = {https://doi.org/10.1016/0010-4655(91)90130-D},
	issn = {0010-4655},
	journal = {Computer Physics Communications},
	number = {3},
	pages = {345-359},
	title = {Feyn Calc - Computer-algebraic calculation of Feynman amplitudes},
	url = {https://www.sciencedirect.com/science/article/pii/001046559190130D},
	volume = {64},
	year = {1991}}

@article{Shtabovenko-2020.11,
	author = {Shtabovenko, Vladyslav and Mertig, Rolf and Orellana, Frederik},
	doi = {10.1016/j.cpc.2020.107478},
	issn = {0010-4655},
	journal = {Computer Physics Communications},
	month = nov,
	pages = {107478},
	publisher = {Elsevier BV},
	title = {FeynCalc 9.3: New features and improvements},
	url = {http://dx.doi.org/10.1016/j.cpc.2020.107478},
	volume = {256},
	year = {2020}}

@article{Patel-2017.09,
	author = {Patel, Hiren H.},
	doi = {10.1016/j.cpc.2017.04.015},
	issn = {0010-4655},
	journal = {Computer Physics Communications},
	month = sep,
	pages = {66--70},
	publisher = {Elsevier BV},
	title = {Package -X 2.0: A Mathematica package for the analytic calculation of one-loop integrals},
	url = {http://dx.doi.org/10.1016/j.cpc.2017.04.015},
	volume = {218},
	year = {2017}}

@article{Lee-2012,
	author = {Lee, Roman},
	month = {12},
	title = {Presenting LiteRed: a tool for the Loop InTEgrals REDuction},
	year = {2012}}

@article{Lee-2014,
	author = {Roman N Lee},
	doi = {10.1088/1742-6596/523/1/012059},
	journal = {Journal of Physics: Conference Series},
	month = {jun},
	number = {1},
	pages = {012059},
	title = {LiteRed 1.4: a powerful tool for reduction of multiloop integrals},
	url = {https://dx.doi.org/10.1088/1742-6596/523/1/012059},
	volume = {523},
	year = {2014}}

@inproceedings{Brau-2003,
	author = {Charles A. Brau},
	title = {Modern Problems in Classical Electrodynamics},
	url = {https://api.semanticscholar.org/CorpusID:117031668},
	year = {2003}}

@article{Guadagnini-2002,
	author = {E. Guadagnini},
	doi = {https://doi.org/10.1016/S0370-2693(02)02811-3},
	issn = {0370-2693},
	journal = {Physics Letters B},
	number = {1},
	pages = {19-23},
	title = {Gravitational deflection of light and helicity asymmetry},
	url = {https://www.sciencedirect.com/science/article/pii/S0370269302028113},
	volume = {548},
	year = {2002}}

@article{Tsukamoto-2017.03,
	author = {Tsukamoto, Naoki},
	doi = {10.1103/physrevd.95.064035},
	issn = {2470-0029},
	journal = {Physical Review D},
	month = mar,
	number = {6},
	publisher = {American Physical Society (APS)},
	title = {Deflection angle in the strong deflection limit in a general asymptotically flat, static, spherically symmetric spacetime},
	url = {http://dx.doi.org/10.1103/PhysRevD.95.064035},
	volume = {95},
	year = {2017}}

@article{Bozza-2001.09,
	author = {Bozza, V. and Capozziello, S. and Iovane, G. and Scarpetta, G.},
	date = {2001/09/01},
	doi = {10.1023/A:1012292927358},
	id = {Bozza2001},
	isbn = {1572-9532},
	journal = {General Relativity and Gravitation},
	number = {9},
	pages = {1535--1548},
	title = {Strong Field Limit of Black Hole Gravitational Lensing},
	url = {https://doi.org/10.1023/A:1012292927358},
	volume = {33},
	year = {2001}}

@article{Frittelli-2000.02,
	author = {Frittelli, Simonetta and Kling, Thomas P. and Newman, Ezra T.},
	doi = {10.1103/PhysRevD.61.064021},
	issue = {6},
	journal = {Phys. Rev. D},
	month = {Feb},
	numpages = {14},
	pages = {064021},
	publisher = {American Physical Society},
	title = {Spacetime perspective of Schwarzschild lensing},
	url = {https://link.aps.org/doi/10.1103/PhysRevD.61.064021},
	volume = {61},
	year = {2000}}

@article{Laporta-2000,
	author = {Laporta},
	doi = {10.1016/s0217-751x(00)00215-7},
	issn = {0217-751X},
	journal = {International Journal of Modern Physics A},
	pages = {5087},
	publisher = {World Scientific Pub Co Pte Lt},
	title = {High-precision calculation of multi-loop Feynman integrals by difference equations},
	url = {http://dx.doi.org/10.1016/S0217-751X(00)00215-7},
	volume = {15},
	year = {2000}}

@misc{Dixon-2014,
	author = {Dixon, Lance J},
	doi = {10.5170/CERN-2014-008.31},
	language = {en},
	publisher = {CERN},
	title = {A brief introduction to modern amplitude methods},
	url = {https://cds.cern.ch/record/1613349},
	year = {2014}}

@article{Bern-2008.10,
	author = {Bern, Z. and Carrasco, J. J. M. and Johansson, H.},
	doi = {10.1103/physrevd.78.085011},
	issn = {1550-2368},
	journal = {Physical Review D},
	month = oct,
	number = {8},
	publisher = {American Physical Society (APS)},
	title = {New relations for gauge-theory amplitudes},
	url = {http://dx.doi.org/10.1103/PhysRevD.78.085011},
	volume = {78},
	year = {2008}}

@article{Bern-2002,
	author = {Bern, Zvi},
	doi = {10.12942/lrr-2002-5},
	issn = {1433-8351},
	journal = {Living Reviews in Relativity},
	month = jul,
	number = {1},
	publisher = {Springer Science and Business Media LLC},
	title = {Perturbative Quantum Gravity and its Relation to Gauge Theory},
	url = {http://dx.doi.org/10.12942/lrr-2002-5},
	volume = {5},
	year = {2002}}

@misc{Brunello-2022,
	archiveprefix = {arXiv},
	author = {Giacomo Brunello},
	eprint = {2211.01321},
	primaryclass = {hep-th},
	title = {Effective Field Theory Approach to General Relativity and Feynman Diagrams for Coalescing Binary Systems},
	url = {https://arxiv.org/abs/2211.01321},
	year = {2022}}

@article{Akhoury-2021.03,
	author = {Akhoury, Ratindranath and Saotome, Ryo and Sterman, George},
	doi = {10.1103/physrevd.103.064036},
	issn = {2470-0029},
	journal = {Physical Review D},
	month = mar,
	number = {6},
	publisher = {American Physical Society (APS)},
	title = {High energy scattering in perturbative quantum gravity at next-to-leading power},
	url = {http://dx.doi.org/10.1103/PhysRevD.103.064036},
	volume = {103},
	year = {2021}}

@article{Bai-2017.03,
	author = {Bai, Dong and Huang, Yue},
	doi = {10.1103/PhysRevD.95.064045},
	issue = {6},
	journal = {Phys. Rev. D},
	month = {Mar},
	numpages = {12},
	pages = {064045},
	publisher = {American Physical Society},
	title = {More on the bending of light in quantum gravity},
	url = {https://link.aps.org/doi/10.1103/PhysRevD.95.064045},
	volume = {95},
	year = {2017}}

@article{Bohr-2015.02,
	author = {Bjerrum-Bohr, N. E. J. and Donoghue, John F. and Holstein, Barry R. and Plant\'e, Ludovic and Vanhove, Pierre},
	doi = {10.1103/PhysRevLett.114.061301},
	issue = {6},
	journal = {Phys. Rev. Lett.},
	month = {Feb},
	numpages = {5},
	pages = {061301},
	publisher = {American Physical Society},
	title = {Bending of Light in Quantum Gravity},
	url = {https://link.aps.org/doi/10.1103/PhysRevLett.114.061301},
	volume = {114},
	year = {2015}}

@misc{Heidari-2024,
	archiveprefix = {arXiv},
	author = {N. Heidari and A. A. Ara{\'u}jo Filho and R. C. Pantig and A. {\"O}vg{\"u}n},
	eprint = {2410.08246},
	primaryclass = {gr-qc},
	title = {Absorption, Scattering, Geodesics, Shadows and Lensing Phenomena of Black Holes in Effective Quantum Gravity},
	url = {https://arxiv.org/abs/2410.08246},
	year = {2024}}

@article{Bohr-2016,
	author = {N. Emil J. Bjerrum-Bohr and John F. Donoghue and Barry R. Holstein and Ludovic Plant{\'e} and Pierre Vanhove},
	journal = {Journal of High Energy Physics},
	title = {Light-like scattering in quantum gravity},
	url = {https://api.semanticscholar.org/CorpusID:7761171},
	volume = {2016},
	year = {2016}}

@article{Gibbons-2008,
	author = {Gibbons, G W and Werner, M C},
	doi = {10.1088/0264-9381/25/23/235009},
	issn = {1361-6382},
	journal = {Classical and Quantum Gravity},
	month = nov,
	number = {23},
	pages = {235009},
	publisher = {IOP Publishing},
	title = {Applications of the Gauss--Bonnet theorem to gravitational lensing},
	url = {http://dx.doi.org/10.1088/0264-9381/25/23/235009},
	volume = {25},
	year = {2008}}

@article{Virbhadra-2000.09,
	author = {Virbhadra, K. S. and Ellis, George F. R.},
	doi = {10.1103/physrevd.62.084003},
	issn = {1089-4918},
	journal = {Physical Review D},
	month = sep,
	number = {8},
	publisher = {American Physical Society (APS)},
	title = {Schwarzschild black hole lensing},
	url = {http://dx.doi.org/10.1103/PhysRevD.62.084003},
	volume = {62},
	year = {2000}}

@article{EHT-2019.04-3,
	author = {Event Horizon Telescope Collaboration et al},
	doi = {10.3847/2041-8213/ab0c96},
	issn = {2041-8213},
	journal = {The Astrophysical Journal Letters},
	month = apr,
	number = {1},
	pages = {L2},
	publisher = {American Astronomical Society},
	title = {First M87 Event Horizon Telescope Results. II. Array and Instrumentation},
	url = {http://dx.doi.org/10.3847/2041-8213/ab0c96},
	volume = {875},
	year = {2019}}

@article{EHT-2019.04-2,
	author = {Akiyama, Kazunori and Alberdi, Antxon and et.al},
	doi = {10.3847/2041-8213/ab0f43},
	issn = {2041-8213},
	journal = {The Astrophysical Journal Letters},
	month = apr,
	number = {1},
	pages = {L5},
	publisher = {American Astronomical Society},
	title = {First M87 Event Horizon Telescope Results. V. Physical Origin of the Asymmetric Ring},
	url = {http://dx.doi.org/10.3847/2041-8213/ab0f43},
	volume = {875},
	year = {2019}}

@article{EHT-2019.04-1,
	author = {Event Horizon Telescope Collaboration et.al},
	doi = {10.3847/2041-8213/ab0e85},
	issn = {2041-8213},
	journal = {The Astrophysical Journal Letters},
	month = apr,
	number = {1},
	pages = {L4},
	publisher = {American Astronomical Society},
	title = {First M87 Event Horizon Telescope Results. IV. Imaging the Central Supermassive Black Hole},
	url = {http://dx.doi.org/10.3847/2041-8213/ab0e85},
	volume = {875},
	year = {2019}}

@article{Donoghue-1994.09,
	author = {Donoghue, John F.},
	doi = {10.1103/physrevd.50.3874},
	issn = {0556-2821},
	journal = {Physical Review D},
	month = sep,
	number = {6},
	pages = {3874--3888},
	publisher = {American Physical Society (APS)},
	title = {General relativity as an effective field theory: The leading quantum corrections},
	url = {http://dx.doi.org/10.1103/PhysRevD.50.3874},
	volume = {50},
	year = {1994}}

@book{Alsing-1998,
	author = {Paul M. Alsing},
	publisher = {Am.J.Phys},
	title = {The optical-mechanical analogy for stationary metrics in general relativity},
	year = {1998}}

@article{R.Keith-Ellis-2008,
	author = {R.Keith Ellis,Giulia Zanderighi},
	journal = {JHEP 02 (2008) 002},
	title = {Scalar one-loop integrals for QCD},
	year = {2008}}

\end{document}